\documentclass[a4paper, 11pt]{llncs}
\usepackage{cmap}
\usepackage[T1]{fontenc}

\usepackage{fullpage}
\usepackage{amsmath, amsfonts, mathtools, amssymb}
\usepackage{hyperref}
\usepackage{algorithm}
\usepackage{algpseudocode}
\usepackage[american]{babel}

\newcommand{\eps}{\varepsilon}
\newcommand{\oeps}{1+\varepsilon}
\newcommand{\ieps}{1/\eps}
\newcommand{\gcbp}{\mathsf{GCBP}}
\newcommand{\bpcc}{\textsf{BPCC}}
\newcommand{\opt}{\mathsf{OPT}}

\newcommand{\items}{\mathsf{T}}
\newcommand{\litems}{\mathsf{L}}
\newcommand{\sitems}{\mathsf{S}}
\newcommand{\cards}{\mathsf{K}}
\newcommand{\class}{\Psi}
\newcommand{\lclass}{\Upsilon}
\newcommand{\sgroups}{\Gamma}
\newcommand{\sparse}{\sigma}
\newcommand{\dense}{\delta}
\newcommand{\bpguess}{\Theta}

\newcommand{\lsizes}{\mathsf{Z}}
\newcommand{\confs}{\mathsf{C}}
\newcommand{\costguess}{\Omega}
\newcommand{\costs}{\Sigma}
\newcommand{\bins}{\mathsf{B}}
\newcommand{\expensivebound}{\phi}

\newcommand{\nre}{\mathbb{Q}_{\geq 0}}
\newcommand{\pre}{\mathbb{Q}_{> 0}}
\newcommand{\nint}{\mathbb{Z}_{\geq 0}}
\newcommand{\pint}{\mathbb{Z}_{> 0}}

\title{APTAS for bin packing with general cost structures}

\title{APTAS for bin packing with general cost structures}
\author{G. Jaykrishnan\textsuperscript{1} \and Asaf Levin \thanks{Supported in part by ISF - Israel Science Foundation grant numbers 1467/22.}\textsuperscript{2}\\
\bigskip
\small
\textsuperscript{1}\texttt{jaykrishnangp@gmail.com},
\textsuperscript{2}\texttt{levinas@technion.ac.il}}
\authorrunning{G. Jaykrishnan \and A. Levin}
\institute{Faculty of Data and Decision Sciences, The Technion, Haifa, Israel}

\begin{document}
\maketitle
\begin{abstract}
We consider the following generalization of the bin packing problem.  We are given a set of items each of which is associated with a rational size in the interval $[0,1]$, and a monotone non-decreasing non-negative cost function $f$ defined over the cardinalities of the subsets of items.  A feasible solution  is a partition of the set of items into bins subject to the constraint that the total size of items in every bin is at most $1$. Unlike bin packing, the goal function is to minimize the total cost of the bins where the cost of a bin is the value of $f$  applied on the cardinality of the subset of items packed into the bin.  We present an APTAS for this strongly NP-hard problem.  We also provide a complete complexity classification of the problem with respect to the choice of $f$.
\end{abstract}

\section{Introduction}

Bin packing is a well-studied problem which has many applications in memory allocation, logistics and production systems.  An input instance to the bin packing problem consists of a set of items each of which has an associated size that is a positive rational number of at most $1$. The goal is, given such an instance, to find a partition of the items into subsets such that the total size of the items in each subset is at most $1$, and the number of subsets is minimized.  Such a subset in a feasible solution is named a bin. Bin packing is a strongly  NP-hard problem and in fact does not admit a polynomial time algorithm whose approximation ratio is strictly better than $\frac 32$ unless P $\neq$ NP.
In  bin packing, the cost of a subset of the partition is $1$ if it is not empty and $0$ if empty.
Here we consider a generalization of the bin packing problem in which the cost of each subset of the partition is a function of the number of items in the subset.
More formally, our problem is defined as follows.

\paragraph{Problem definition.}
The problem that we study here is called the \textsc{Bin packing with general cost structures} denoted by $\gcbp$. The input instance of $\gcbp$ consists of a set of items $\items = \{1,2,\ldots,n\}$ where each item $i\in\items$ has a size $s_i\in[0,1]$, and a cost function $f:\{0,1,2,\ldots,n\}\rightarrow \nre$ which acts on the cardinality of a bin to give the cost of the bin. We assume $f$ is a monotonically non-decreasing function with $f(0)=0$. Furthermore, we scale $f$ so that $f(1)=1$. A feasible solution of $\gcbp$ is a partition of $\items$ into some $k$ number of subsets $\bins_1,\bins_2,\dots,\bins_k$ such that $\sum_{i\in\bins_j} s_i \leq 1$ for $j=1,2,\ldots, k$ where $k$ is chosen by the algorithm and different solutions may use different values of $k$. The cost of a feasible solution is defined as the sum of the costs of each subset that forms the partition of $\items$ in the feasible solution i.e., $\sum_{j=1}^{k}f(|\bins_j|)$. The goal of $\gcbp$ is to find a feasible solution such that the cost of the solution is minimized. $\gcbp$ is strongly NP-hard since it generalizes bin packing. We develop an APTAS for $\gcbp$ in this paper.  Some applications of this problem were described in \cite{anily1994worst,epstein2012bin}.  In particular, applications in cryptography, quality control, reliability, and vehicle routing are described in \cite{anily1994worst}, where an application in a multiprocessor system is given in  \cite{epstein2012bin}. Note that in our problem we allow zero-sized items and their existence may modify the optimal solutions of the items of non-zero size.

\paragraph{Definitions of approximation schemes.}  Before getting to the literature review, we define the metric that we study for analyzing the algorithms.  Following the studies of bin packing algorithms we use the asymptotic approximation ratio measure.
We define the \emph{asymptotic approximation ratio} of an algorithm $\mathcal{A}$ as the infimum $\mathcal{R} \geq 1$ such that for any input, $C(\mathcal{A}) \leq \mathcal{R}\cdot\mathsf{OPT}+c$, where $C(\mathcal{A})$ is the cost of the output of $\mathcal{A}$ and $\mathsf{OPT}$ is the cost of an optimal solution for the same input, and $c$ is a constant independent of the input. An \emph{asymptotic polynomial time approximation scheme} is a family of approximation algorithms such that for every $\eps >0$, the family contains a polynomial time algorithm with an asymptotic approximation ratio of $\oeps$. We abbreviate asymptotic polynomial time approximation scheme by APTAS (also called an asymptotic PTAS). An \emph{asymptotic fully polynomial time approximation scheme} (AFPTAS) is an APTAS whose time complexity is polynomial not only in the length of the binary encoding of the input but also in $1/\eps$.  See \cite{vazirani2001approximation} for an exposition of these definitions and related results.

\paragraph{Literature review and related problems.}
We believe we are the first to deal with $\gcbp$ as defined above.  Beside bin packing, that is a special case of $\gcbp$ as explained above, there are studies of two other special cases of the problem.  The first we consider below is $\gcbp$ under the restriction that the function $f$ is a concave function (referred below as $\gcbp$ with concave cost).  The second special case we consider here is the {\em bin packing with cardinality constraint} (\bpcc).

The $\gcbp$ with concave cost was first studied in \cite{anily1994worst}. They discuss how some heuristics for classical bin packing fare when used on this problem from a worst-case perspective. \cite{bramel1997average} treated this problem from an average-case perspective and developed techniques to determine the asymptotic optimal solution value.  This type of study clearly works only for special cases of the cost functions as for general functions there is no finite bound applied to all functions, so the algorithm needs to consider the specific cost function. The results of  \cite{anily1994worst,bramel1997average} do not use the definition of $f$ in order to modify the algorithms or their analysis, so the bounds apply uniformly for all functions $f$ simultaneously.  In order to prove much better bounds, and in fact to obtain an AFPTAS, \cite{epstein2012bin} considered the function $f$ as a part of the input and their scheme uses the values of $f$ for creating the solution. The techniques they used to pack the small items fail to work for $\gcbp$ because they leverage the concavity of the cost function to develop an auxiliary packing algorithm for the small items.

\bpcc{} limits the number of items in each bin to be at most an integer $K\in(1,n)$, and it is a special case of $\gcbp$ with a cost function
\begin{align*}
    f(c) =
    \begin{cases}
        1, & c \in [1,K]       \\
        x, & c \in [K+1, n]  ,
    \end{cases}
\end{align*}
where $x$ is a relatively large value like $n$.
\cite{kellerer1999cardinality} presents an approximation algorithm, \cite{caprara2003approximation} presents an APTAS, and \cite{epstein2010afptas,jansen2019approximation,jaykrishnan2024scheduling} present AFPTASs for \bpcc{}. A conventional technique to deal with \bpcc{} is to consider two cases of the problem: when $K$ is small and when $K$ is large. Each case is then considered independently and schemes are established for each. This technique cannot be used in $\gcbp$ since a feasible solution may contain bins of both large and small cardinality. We will have implied cardinality constraints on the bins in $\gcbp$ but different bins may have very different implied cardinalities. The classical bin packing problem is a special case of $\gcbp$ obtained by setting the cost function to be $f(c)=1,\forall c\in[1,n]$ in $\gcbp$. An AFPTAS for the bin packing problem is known since the seminal works of \cite{fernandez1981bin,karmarkar1982efficient}.

We conclude this literature review by mentioning that there is another generalization of bin packing called bin packing with bin utilization cost.
This problem differs from $\gcbp$ in the fact that the cost function is a non-decreasing function of the total size of items in a bin or bin utilization and not of its cardinality.  We refer to \cite{epstein2017afptas} for an AFPTAS for this problem, and \cite{leung2008asymptotic,li2006bin,haouari2024lower} to some other results for special cases of this variant. The techniques used for bin packing with bin utilization cost clearly do not carry over to $\gcbp$.

\paragraph{Our results and paper overview.}
We start our study by presenting a complete complexity classification of $\gcbp$ in terms of the function $f$ in Section \ref{sec:class}. We provide an easily tested criterion that determines for a given $f$ if the problem is NP-hard or polynomial time solvable.  Our main result, that is, an APTAS for $\gcbp$ is described in the later sections of our work.  As mentioned above our problem generalizes the \bpcc{} problem, but we cannot use the methodology of partitioning the instances into small cardinality bound instances and large cardinality bound instances.  This impossibility arises as clearly there are instances for $\gcbp$ for which there are many bins with small cardinality and a major part of the cost incurred by a small number of bins with very high cardinality.  We provide some highlight structure of the scheme in Section \ref{sec:over}, and the scheme itself in Section \ref{first-stage} and Section \ref{second-stage}.

\paragraph{Notation.}
We let $\nre$ denote the set of non-negative rational numbers, $\pre$ the set of positive rational numbers, $\nint$ the set of non-negative integers, and $\pint$ the set of positive integers.  For a $c\in\pint$, $[c]$ denotes the set of positive integers up to and including $c$, i.e., $[c]=\{1,\ldots, c\}$.  We fix a value of $\eps>0$ such that $\frac{1}{\eps}$ is an integer.

\section{Complexity classification of $\gcbp$\label{sec:class}}
In this section we consider the following questions.  Let $f$ be a given cost function that is monotone non-decreasing with $f(0)=0$ and $f(1)=1$.  Under what conditions on $f$, the corresponding restriction of $\gcbp$ is polynomial time solvable, and can we prove that in the remaining cases, the problem is NP-hard in the strong sense?  We fully answer this question, by considering the following allocation of cost to items.

For $j\in[n]$, let $F(j) = \frac{f(j)}{j}$. $F(j)$ is the average cost of an item packed into a bin with a subset of items of cardinality $j$. The cost of a packing can then be written as $\sum_{i\in\items} F(|b(i)|)$ where $b(i)$ is the bin in which item $i$ is packed. We would like to consider the set $S=\text{argmin}_j F(j)$ of the minimizers of $F$, and let $k$ be the minimum value in $S$, i.e., the minimizer of $F$ breaking ties in favor of a smaller value.  That is, $k = \min_{\kappa} \{ \kappa\in S\}$.   In this section we prove that if $k=1$ or $k=2$, the problem is polynomial time solvable, whereas if $k\geq 3$, it is NP-hard in the strong sense.  We consider first the positive results.

\begin{theorem}\label{theorem-1}
    If $k=1$, then $\gcbp$ is polynomial time solvable.
\end{theorem}
\begin{proof}
    We argue that in this case there is an optimal packing that packs each item in a dedicated bin. Assume that the claim does not hold, then consider an optimal packing in which the number of bins whose cardinality is strictly greater than $1$ is minimized.  Then by assumption it has a bin $B$ with more than one item packed in $B$. We modify this solution by packing the items packed in $B$ in dedicated bins. The cost of the items that used to be packed in $B$ in the new solution is at most the cost in the old solution since $k=1$. Thus, either the cost of the new packing is less than the considered optimal solution or the number of bins with cardinality larger than $1$ in the new solution is smaller than in the old solution.  In both cases it is a contradiction.
    \qed\end{proof}

\begin{theorem}
    If $k=2$, then $\gcbp$ is polynomial time solvable.
\end{theorem}
\begin{proof}
    In this case we prove the following structure of an optimal solution. We argue that there is an optimal packing that has at most one bin with an odd cardinality at least $3$ and all other bins in this optimal solution have cardinality $1$ or $2$.  Consider a fixed optimal solution $\opt$ such that among all optimal solutions $\opt$ minimizes the number of bins with even cardinality larger than $2$, and among all optimal solutions that minimize this number of bins with even cardinality larger than $2$, we select $\opt$ to be one that minimizes the number of bins with odd cardinality of at least $3$.

    First assume that there is a bin $B$ in $\opt$ of even cardinality $x$ of at least $4$. Then, we replace $B$ with a set of bins packing the items that used to be packed in $B$ in pairs. Since $k=2$, $f(x)\geq \frac{x}{2} \cdot f(2)$ and the cost of the new solution is not larger than the cost of $\opt$ while the number of even cardinality bins of cardinality at least $4$ has decreased.  This is a contradiction to the choice of $\opt$.  Thus, $\opt$ does not have a bin $B$ whose cardinality is even and at least $4$.
    Next, assume by contradiction that $\opt$ has at least two bins $B$ and $B'$ each with odd cardinality at least $3$. Since the cardinality of the item set of each of these bins ($B$ and $B'$) is larger than $2$, the smallest item in each such bin has size at most $\frac 12$, so we can pack a smallest item of $B$ together with a smallest item of $B'$ into a common bin.  The other items of these bins are also packed in pairs so that each pair contains two items that used to be packed in a common bin.  Denote by $x$ and $x'$ the cardinalities of $B$ and $B'$ in $\opt$, respectively.  The new solution has $\frac{x+x'}{2}$ bins that are used to pack the items from $B$ and $B'$ with total cost $\frac{x+x'}{2}\cdot f(2)  \leq f(x)+f(x')$ where the inequality holds by the definition of $k$ and the assumption $x,x'\geq 2$. This is a contradiction to the choice of $\opt$.   This concludes the structure claim on $\opt$.

    Next, we present the algorithm to solve $\gcbp$ in this case.
    First, we guess the number of bins in $\opt$ with cardinality equal to $1$, and the cardinality of the bin with odd cardinality of at least $3$ (if it exists). There is a linear number of guesses for each. Denote the guessed number of bins with cardinality equal to $1$ with $\rho_s$, and denote the guessed cardinality of the bin with odd cardinality at least $3$ by $\rho_{\ell}$. If the packing does not contain a bin of odd cardinality of at least $3$, then $\rho_{\ell}=0$.  We let $\rho_p=n-\rho_s-\rho_{\ell}$ be the number of items that are packed in pairs in $\opt$.  By guessing we mean that we enumerate all possibilities for these parameters, for each possibility we test in polynomial time if there is a feasible solution satisfying these parameters.  For a guess, we have that all feasible solutions satisfying these guessed values have a common cost of $\rho_s \cdot f(1) + f(\rho_{\ell})+ \frac{\rho_p}{2} \cdot f(2)$.  Thus, we can output a feasible solution of minimum cost in polynomial time.

    We next consider a specific triple $(\rho_s,\rho_p,\rho_{\ell})$ and assume that there exists a feasible solution with these values.  We first note that without loss of generality we can assume that the $\rho_s$ items packed in dedicated bins (one per bin) are the largest items in the instance.  This observation follows by a trivial exchange argument.  Thus, we remove the $\rho_s$ largest items from the instance, and then we check if the remaining items can be packed into $\rho_p$ pairs and one bin with $\rho_{\ell}$ items.  To do that we look for the collection of $\rho_p$ pairs (of the remaining items) so that the total size of items in these pairs is maximized and each pair of items has total size at most $1$.  Such collection of pairs can be found in polynomial time by searching for a maximum weight matching of cardinality $\frac{\rho_p}{2}$ in a graph whose node set is the remaining items and two nodes corresponding to items $i_1,i_2$ have an edge between them if $s_{i_1}+s_{i_2}\leq 1$ and if so, its weight is $s_{i_1}+s_{i_2}$.  After finding an optimal matching in this graph we check if the total size of items corresponding to isolated nodes is at most $1$.  We have that if the total size of these isolated nodes is at most $1$, then these items fit into a common bin, so there is a feasible solution corresponding to the current guess.  If on the other hand, their total size is strictly larger than $1$, then no matter what is the selected set of $\frac{\rho_p}{2}$ pairs, the total size of remaining items is strictly larger than $1$, and we can safely conclude that the current guess does not correspond to a feasible solution.
    \qed
\end{proof}

We next show that for every fixed value of $k$ that is at least $3$, the problem $\gcbp$ is NP-hard in the strong sense.  To make a precise statement of this claim, denote by $\gcbp (k)$ the subset of the instances for the problem $\gcbp$ where the function $f$ has the corresponding value of $k$.  This subset is defined for every value of $k$, and we have seen above that if $k=1$ or $k=2$, these are polynomial time solvable subsets of instances.  Next, we consider the corresponding restriction for higher values of $k$.

\begin{theorem}
    Let $k\geq 3$ be a fixed constant.  Then, $\gcbp (k)$ is strongly NP-hard.
\end{theorem}
\begin{proof}
    We prove this via reduction from 3-Partition to $\gcbp (k)$.
    An input instance to 3-Partition is a set $I=\{s_1,s_2,\ldots,s_{3m}\}$ of positive integers, a positive integer bound $Z$, such that each of the $3m$ integers are in $(Z/4,Z/2)$ and $\sum_{i=1}^{3m} s_i =mZ$. A feasible solution to 3-Partition is a partition of $I$ into $m$ subsets $S_1,S_2,\ldots,S_m$ such that for each subset $S_j$, we have $\sum_{i\in S_j} s_i = Z$.  In particular, it means that each such subset has exactly three elements. We transform this instance to an instance of $\gcbp (k)$. The input instance is a set of items $I'=I\cup A$ where $A$ is a set of zero-sized items such that $|A| = m(k-3)$. The size of an item $i\in I$ is scaled to $\frac{s_i}{Z}$. A feasible solution of $\gcbp$ is a partition of $I'$ into some $j$ number of subsets (bins) such that the total size of the items in a bin is at most $1$. That is, the total size of items in every bin before the scaling by $Z$ is at most $Z$. The transformation requires only linear time. Notice that at most $3$ non-zero sized items (i.e., from $I$) can be contained in any bin.

    We next claim that the optimal solution cost of $\gcbp (k)$ is at most $f(k) \cdot m$ if and only if the 3-Partition instance is a YES instance.  First assume that the Partition instance is a YES instance.  In order to create a corresponding solution for $\gcbp (k)$ we add $k-3$ zero sized items to every subset $S_i$ in the feasible solution for the 3-Partition instance.  Then, we get $m$ subsets, each of which has $k$ items of total scaled size of $1$, so the resulting solution is a feasible solution to the $\gcbp (k)$ instance of cost $f(k) \cdot m$.

    Next, assume that the 3-Partition instance is a NO instance.  Then, in every feasible solution to the $\gcbp (k)$ instance there must be at least one bin in which the number of packed items is strictly smaller than $k$ (but at least $1$).  We partition the cost of an optimal solution $\opt$ to the $\gcbp (k)$ instance among the items, so that if an item is packed into a bin with cardinality $i$, then it is assigned a cost of $\frac{f(i)}{i}$.  Then the total assigned cost of the items is exactly the cost $\opt$.  However, if the cardinality of a bin is strictly smaller than $k$, then items packed there have assigned cost strictly larger than $\frac{f(k)}{k}$ while if an item is packed into a bin of cardinality at least $k$, then its assigned cost is not smaller than   $\frac{f(k)}{k}$.  Since $\opt$ has at least one item packed into a bin with cardinality strictly smaller than $k$, the total assigned cost (that equals the cost of $\opt$) is strictly larger than $f(k) \cdot m$.
    \qed
\end{proof}

\section{The APTAS: preliminaries and overview \label{sec:over}}

The optimal solution of $\gcbp$ is denoted by $\opt$.  We extend the definition of $f$ to include a value of a collection $S$ of bins so that its value $f(S)$ is $f(S)=\sum_{B\in S} f(|B|)$. Thus, the cost of $\opt$ is $f(\opt)=\sum_{j=1}^{k}f(|\bins_j|)$ where $\{\bins_j:j\in[k]\}$ is the partition of $\items$ in $\opt$.
A bin is \emph{sparse} if the number of items in the bin is in $[1,\ieps^2]$. A bin $b$ is \emph{dense} if the number of items in $b$ is strictly greater than $\ieps^2$. If we fix a feasible solution to $\gcbp$, then we have a partition of the item set into the items that are packed into sparse bins, and the items that are packed into dense bins.  We let $\opt_\dense$ be the collection of dense bins and $\opt_\sparse$ be the collection of sparse bins in $\opt$.

In a sense we would like to guess the set of items packed into $\opt_\dense$ (and the remaining items are packed into $\opt_\sparse$).  Note that we are not able to do so with a polynomial number of guesses.  Thus, we relax this requirement using a delicate combinatorial argument. We show that there exists a subset of the feasible solutions containing a near optimal solution so that this subset of feasible solutions has only polynomial number of such partitions of the item set. Then we can enumerate only on those partitions of the item set. This novel approximate partition is our main contribution. Then, we approximate each such sub-instance independent of the other sub-instance. We use the conventional approaches of linear grouping \cite{fernandez1981bin}, the use of a mixed integer linear program or an integer program to find a succinct description of a feasible solution with cost close to $\opt$, and rounding the cost function \cite{epstein2012bin}.

The first stage (Section \ref{first-stage}) of the scheme deals with this partition of the item set that is the main goal of the preprocessing step, and  approximates the packing of the sparse bins of $\opt$.  The approximation of the packing of the sparse bins is based on using standard techniques of linear grouping and a configuration integer program of fixed dimension. The second stage (Section \ref{second-stage}) approximates the packing of the dense bins of $\opt$. Since we already guessed the items packed into the sparse bins, we have the remaining items that $\opt$ packs into dense bins. We can use the usual techniques to build the remaining components of the scheme but here we need to be careful regarding a situation where there are very few bins with very large cardinalities and thus incur a major part of the cost of $\opt_\dense$.

\section{The first stage}\label{first-stage}
The first stage starts with guessing the item set of the sparse bins in a near optimal solution. This guessed set of items will be named the {\em sparse instance} while the complement item set will be called the {\em dense instance}. The sparse instance is later rounded to restrict the item sizes in the instance by applying linear grouping on its entire item set. In a later step of this stage, we define configurations based on this rounded instance and use them in an integer program to find the packing that approximates $f(\opt_\sparse)$.  The motivation to the partitioning step we consider next is the fact that we intend to apply linear grouping of the item set of the sparse instance.

First, we group the items in the input instance into at most $\ieps^3$ groups denoted by $\sgroups_i, i\in[\ieps^3]\cup \{0\}$. The items of the input instance are grouped using the following condition -- the items of group $\sgroups_i$ are the largest items of  $\items\backslash\bigcup_{j=0}^{i-1}\sgroups_j$ (breaking ties in favor of smaller indexes).  We stress that unlike linear grouping we do not impose cardinality conditions on these groups. These groups will be used to provide structure to guess the sparse instance.

Our scheme will first guess the groups of the input instance. This is done by guessing the largest item size in each group.  The guess is encoded as a vector of length $\ieps^3+1$ named the \emph{breakpoint guess}. The value of the $(i+1)^{th}$ component of the breakpoint guess will be the value of the largest item size in $\sgroups_i$ (together with the index of this item breaking ties in favor of smaller index) and a zero value for any component $i+1$ means that $\sgroups_i$ is empty. A breakpoint guess is feasible if the non-zero breakpoints are in non-increasing order. Each component can have $n+1$ possible values, thus there are $(n+1)^{(\ieps^3+1)}$ possible breakpoint guesses (a polynomial number of guesses once $\eps$ is fixed). Let $\bpguess$ be the feasible breakpoint guess.

Once we have $\bpguess$, the groups can be defined as follows. Group $\sgroups_i$ is the set of all items that have sizes at most $\bpguess_{i+1}$ and greater than $\bpguess_{i+2}$ (breaking ties according to index) where $\bpguess_{\ieps^3+2}$ is defined as the smallest item size in the instance. We use group $\sgroups_0$ as an auxiliary group to allow us to pack the largest items from the input instance into  dense bins.

\subsubsection{Encoding a partition of the item set into sparse instance and dense instance.}  A partition of a family of solutions that contains at least one near optimal solution is defined according to the following rules where we algorithmically enumerate over all breakpoint guesses and over all possible numbers of items in the sparse instance. Let $\Pi$ be the sparse instance cardinality guess and let $\bpguess$ be the breakpoint guess.

For each $i\in[\ieps^3]$, only the largest items from group $\sgroups_i$ will be packed in the sparse bins and that subset is called a \emph{class} and denoted by $\class_i$ and furthermore the classes should satisfy the following properties.
\begin{enumerate}
    \item $\left\lceil \eps^3\Pi \right\rceil = |\class_1| \geq |\class_2| \geq \ldots \geq |\class_{\ieps^3-1}| \geq |\class_{\ieps^3}| = \left\lfloor \eps^3\Pi \right\rfloor$, and \label{sparse-property-1}
    \item items of class $\class_i$ are the largest items in $\sgroups_i$.\label{sparse-property-2}
\end{enumerate}
Given this partition we further round up the sizes of items in each class $\class_i$ so that the rounded size of every item in class $\class_i$ is defined as the largest size of an item of the class.  That is, for every $i$, the rounded size of each item of $\class_i$ is $\bpguess_{i+1}$.  If an item does not belong to any of the classes it will be part of the dense instance, and we do not modify its size.
The sparse instance is $\bigcup_{i=1}^{\ieps^3}\class_i$ and $|\bigcup_{i=1}^{\ieps^3}\class_i|=\Pi$ where the items of the sparse instance have rounded size.

The sparse properties, the sparse instance cardinality guess, and the breakpoint guess uniquely define the classes and consequently the sparse instance. Let the number of items in the classes in the guessed sparse instance be denoted by $\Pi_i,i\in[\ieps^3]$. Some classes could be empty depending on $\Pi$.

\begin{lemma}\label{sparse_grouping}
    There is a partition that we test in one of the iterations for which there is a feasible solution whose cost is at most $ (\oeps)f(\opt) + 1$ such that this feasible solution uses (only) dense bins to pack the dense instance and (only) sparse bins to pack the sparse instance.
\end{lemma}
\begin{proof}
    Consider the optimal packing $\opt$ with cost $f(\opt)$. We will transform $\opt$ into another packing $\opt'$ that satisfies the claimed properties. Recall that the collection of sparse bins of $\opt$ is $\opt_{\sparse}$ and the collection of dense bins is $\opt_{\dense}$, and $f(\opt_{\sparse})$ is the cost of the sparse bins and $f(\opt_{\dense})$ is the cost of the dense bins.

    Now we are going to define the iteration of the exhaustive search, namely $\bpguess$ and $\Pi$.  $\Pi$ is simply the number of items that $\opt$ packs in sparse bins.  Based on this value of the guess together with the property \ref{sparse-property-1}, we identify the cardinalities of the classes that we would like to obtain.  We sort the input instance in a non-increasing order of size breaking ties in favor of index.  For every $i$, let $\tau_i$ be the total number of items in the first $i-1$ classes.  Observe that based on $\Pi$, these values are well-defined.  Then, $\bpguess_0=1$, and $\bpguess_i$ is the size of the $(\tau_i+1)$-th largest item in the collection of sparse bins in $\opt$ (together with its index).

    Based on these values of $\bpguess$ and $\Pi$, we define the classes and the rounded size of the items in the sparse instance. We apply the following modifications to the packing of $\opt$.  First, for every group $\sgroups_i$ we apply the following changes on its items. Whenever a dense bin has an item of $\class_i$, we replace it by an item of $\sgroups_i\setminus \class_i$ that is packed into a sparse bin.  This transformation may violate the total size constraints on the bins of $\opt_{\sparse}$, but every dense bin remains a feasible bin with the same cardinality.

    Next, we consider the temporary (infeasible) solution we have. We modify the packing of its sparse bins so that it will be feasible also with respect to the rounded size.  For that, we apply the standard reasoning of linear grouping.  Each item of class $\class_1$ is packed into a dedicated bin.  These dedicated bins are sparse bins, and their total cost satisfies $|\class_1|\leq \eps^3 \Pi + 1$. From the definition of sparse bins and monotonicity of the cost function, $\eps^3\Pi + 1 \leq \eps^3 \cdot \ieps^2 \cdot |\opt_\sparse| + 1 = \eps |\opt_\sparse| + 1 \leq \eps f(\opt_\sparse) + 1$.  Each item of $\class_i$ (for $i\geq 2$) is packed in a position of an item of $\sgroups_{i-1}$ in the sparse bins of $\opt$.  We have that the size of the item of $\sgroups_{i-1}$ is not smaller than the size of the largest item of $\class_i$. That is, the rounded size of the item of class $i$ that enters the bin is at most the size of the item that leaves it. After applying these changes for all classes, we get a feasible solution where the cardinality of every bin that used to be a sparse bin (in $\opt$) is not larger than its cardinality in $\opt$.  The feasibility of this step follows by the fact that the number of items of $\sgroups_{i-1}$ that used to be packed in sparse bins is not smaller than the number of items of $\class_i$.  By the monotonicity of the cost function, we conclude that the resulting solution denoted as $\opt'$, costs at most  $ (\oeps)f(\opt) + 1$ and it satisfies the required properties in the statement of the lemma.
    \qed\end{proof}

Let the optimal solution of the (rounded) sparse instance be denoted by $\opt''_\sparse$.

\subsubsection{Packing the sparse instance using a configuration integer program.}

We use an IP to find a packing of the item set $\bigcup_{i=1}^{\ieps^3}\class_i$ that is close to the optimal packing (with respect to the cost function).
We use configurations to help pack the rounded sparse instance in sparse bins. A \emph{configuration} represents the packing of a bin. It is a vector of length $\ieps^3$. A configuration is denoted by $c$ and a component $c_{i}$ denotes the number of items of class $\class_{i}$ of the rounded instance packed in a sparse bin assigned configuration $c$. A configuration is feasible if the total number of items packed in a bin based on that configuration is at most $\ieps^2$ i.e., $\sum_{i\in[\ieps^3]}c_{i} \leq \ieps^2$, and the total size of its items is at most $1$. Note that the first constraint holds because we consider only sparse bins. The cost of a configuration $c$ is denoted by $f(c)$ and is defined as $f(c) = f\left(\sum_{i\in[\ieps^3]}c_i\right)$. Let $\confs$ be the set of feasible configurations.

\begin{lemma}
    The number of feasible configurations is at most a constant once $\eps$ is fixed.
\end{lemma}
\begin{proof}
    From the feasibility condition for the configurations, the value of each component of the configuration can be at most $\ieps^2$. Thus, the number of possible values for each component is $\ieps^2+1$. Thus, the number of feasible configurations is at most $(\ieps^2+1)^{(\ieps^3)}$ and when $\eps$ is fixed it is at most a constant.
    \qed\end{proof}

\paragraph{The Integer Program.}
We use an integer program (IP) to pack the guessed sparse instance. The decision variables are $y_{c},\forall c\in\confs$, and a variable $y_c$ denotes the number of sparse bins packed based on configuration $c$. We have a constant number of decision variables once $\eps$ is fixed. The IP is as follows.
\begin{align}
    \min\quad
     & \sum_{c\in\confs}f(c) y_c                                                                   \\
    \text{s.t.}\quad
     & \sum_{c\in\confs}c_{i} y_c = \Pi_i, \forall i\in[\ieps^3] \label{IP_cardinality_constraint} \\
     & y_c \in\nint, \forall c\in\confs
\end{align}

\begin{lemma}\label{final_sparse_cost}
    The cost of the optimal solution of the IP is at most $(1+\eps)\cdot f(\opt_\sparse)+1$.
\end{lemma}
\begin{proof}
    We generate a solution to the IP from $\opt''_\sparse$ and the cost of the generated solution gives the upper bound on the cost of an optimal solution to IP.
    For each sparse bin, generate a configuration by identifying the number of items of each class in the bin. This gives us a multi-set of all the configurations corresponding to the bins in $\opt''_\sparse$. For each $c\in\confs$, let $y_c$ be the multiplicity of configuration $c$ in the multi-set. This gives us the solution to the IP. Observe that since every item of the sparse instance is packed in $\opt''_\sigma$, we conclude that constraint \eqref{IP_cardinality_constraint} are satisfied. So indeed we have defined a feasible solution to the IP.

    By the rule we use to generate a configuration for each bin, the number of items in the configuration and the corresponding bin are equal. Thus, the cost of the optimal solution of the IP will be at most the cost of $\opt''_\sparse$ which is at most $(1+\eps)\cdot f(\opt_\sparse)+1$.
    \qed\end{proof}

We solve the IP using Lenstra's algorithm \cite{lenstra1983integer}. We create a packing from the optimal solution $\textbf{y}^*$ of the IP. For each $c\in\confs$ in the support of $\textbf{y}^*$, do the following. Assign $y^*_c$ bins with configuration $c$. For each $i\in[\ieps^3]$, place $c_{i}$ items of class $\class_i$ in each such bin. This is feasible because $\textbf{y}^*$ and the configurations are feasible.

\section{The second stage}\label{second-stage}
In this stage we pack the \emph{dense instance} which is $\items\backslash\bigcup_{i=1}^{n}\class_i$. In this second stage we use many traditional methods for developing approximation schemes for bin packing problems. We divide the item set into small and large, and then restrict the item sizes in the set of large items using linear grouping. The cost function is also rounded to enable us to absorb the cost of many relatively cheap bins using the cost of the most expensive bin used by an optimal solution (that we guess).   These steps are performed in the preprocessing stage. The resulting rounded instance is then packed using a mixed-integer linear program to get a packing close to the optimal packing of the dense bins with respect to the cost function.

Note that we have already used a guessing step in the first stage to get the sparse instance. This guessing makes our scheme an APTAS and not an AFPTAS. Thus, we allow ourselves to use a mixed-integer linear program (MILP) instead of a linear program (LP) in this stage to decrease the number of technical details involved. We similarly used an IP and not LP in the first stage.

\subsection{Preprocessing}\label{second-stage-preprocessing}

\subsubsection{Linear grouping and rounding of large items.}
In what follows we consider the item set  $\items\backslash\bigcup_{i=1}^{n}\class_i$ and ignore the other items.
An item is \emph{large} if its size is at least $\eps$ else it is \emph{small}. The set of large items is denoted by $\litems$ and the set of small items is denoted by $\sitems$. Linear grouping is performed only on the large items. $\litems$ is linearly grouped into $\ieps^3$ classes, denoted by $\lclass_{i},i\in[\ieps^3]$, satisfying the following conditions. The items of class $\lclass_i$ are the largest items in $\litems\backslash\bigcup_{j=1}^{i-1} \lclass_{j}$ and $\left\lceil \eps^3|\litems| \right\rceil = |\lclass_1| \geq |\lclass_2| \geq \ldots \geq |\lclass_{\ieps^3}| = \left\lfloor \eps^3|\litems| \right\rfloor$. Class $\lclass_1$ is removed and packed in dedicated bins (one item per bin).

\begin{lemma}\label{class1_cost}
    The cost of packing the items of class $\lclass_1$ is at most $\eps f(\opt_\dense) + 1$.
\end{lemma}
\begin{proof}
    The number of items in class $\lclass_1$ is $|\lclass_1| \leq \eps^3|L| + 1$ by the properties of linear grouping. Since the cost function is monotone non-decreasing and $f(1)=1$, $|\opt_\dense| \leq f(\opt_\dense)$. Furthermore, since the large items have size at least $\eps$, $\eps|\litems| \leq \sum_{i\in\litems}s_i \leq |\opt_\dense|$. The last two inequalities imply that the cost of packing each item of $\lclass_1$ in a dedicated bin, namely, $|\lclass_1|$ satisfies that, $|\lclass_1| \leq \eps f(\opt_\dense) + 1$.
    \qed\end{proof}

The items in the remaining classes are rounded up to the size of the largest item in the class to which it belongs to. The items of classes $\lclass_i, i\in[\ieps^3]\backslash \{1\}$ (with their rounded size) along with the small items (with their original size that is also named the rounded size) constitute the \emph{rounded instance}. The distinct item sizes of large items in the rounded instance is denoted by $\lsizes$ and $|\lsizes| \leq \ieps^3-1$.

\begin{lemma}\label{dense_rounded_instance}
    The cost of the optimal packing of the rounded instance is at most $f(\opt_\dense)$.
\end{lemma}
\begin{proof}
    We generate a packing for the rounded instance from $\opt_\dense$ and denote it by $\opt'_\dense$. The number of bins in $\opt'_\dense$ will be at most the number of bins in $\opt_\dense$. The packing of small items in $\opt'_\dense$ will be identical to the packing of small items in $\opt_\dense$. For each $i\in[\ieps^3]\backslash\{1\}$ do the following. Place the rounded items from $\lclass_i$ in $\opt'_\dense$ into the positions of the original items of $\lclass_{i-1}$ in $\opt_\dense$ (one item per position). From the properties of linear grouping, this procedure is feasible and does not increase the number of items in any bin. Thus, the cost of each bin in $\opt'_\dense$ is at most the cost of the corresponding bin in $\opt_\dense$. That is, the cost of $\opt'_\dense$ is at most $f(\opt_\dense)$.
    \qed\end{proof}

\subsubsection{Rounding the cost function}

There are at most $n$ distinct values of the cost function. Each distinct value of the cost function is rounded up to the next integer power of $\oeps$. Let $\costs'$ be the set of distinct values for the rounded cost function. For each distinct rounded value, find the maximum cardinality corresponding to that value and let $\cards'$ be the set of all such cardinalities. Note that $|\costs'| \leq \log_{(\oeps)}f(n) + 1$. The rounded cost function is denoted by $g(\cdot)$. For every bin $b$ in $\opt_\dense$, there exists some value $k\in\cards'$ such that $g(|b|)=g(k)$ and $|b|\leq k$. Hence, we can restrict our attention to cardinalities in $\cards'$ once we refer to such cardinality as an upper bound on the cardinality of bins. The inverse of the rounded cost function is denoted by $g^{-1}(\cdot)$ and is defined as follows. For some $y\in\Sigma'$, the inverse of $y$ is the value $k\in\cards'$ such that $g(k)=y$ i.e. $g^{-1}(y)=k$.

\begin{lemma}\label{rounded_cost}
    The total cost of bins in $\opt_\dense$ with respect to the rounded cost function is at most $(\oeps) \cdot f(\opt_\dense)$.
\end{lemma}
\begin{proof}
    $f(\opt_\dense)$ is based on the original cost function, and let $g(\opt_\dense)$ be the cost of the dense bins in $\opt$ based on the rounded cost function. From the definition of this rounded cost, we get $f(k) \leq g(k) \leq (\oeps) \cdot f(k), \forall k\in[n]$. Thus,
    \begin{align*}
        g(\opt_\dense) = \sum_{b\in \opt_\dense} g(|b|) \leq \sum_{b\in \opt_\dense} (\oeps)f(|b|) = (\oeps)f(\opt_\dense) \ .
    \end{align*}
    \qed\end{proof}

\subsubsection{Guessing}

We guess the following information of $\opt_\dense$.
The maximum rounded cost of a bin in the optimal solution is guessed from $\costs'$ and denoted by $\costguess$.
In the analysis we assume that $\costguess$ is indeed the maximum rounded cost of a bin in $\opt_{\dense}$, but algorithmically we apply the algorithm below for every value of the guess, and among the iterations for which we output a feasible solution to $\gcbp$, we output the cheapest one.

\subsection{Packing the dense instance}\label{second-stage-packing}
Items of the rounded instance are packed using a MILP with the aid of configurations.

\subsubsection{Configurations}

A \emph{configuration} is a vector of length $|\lsizes|+1$ and represents the packing of a bin. The components of a configuration $c$ are as follows.
Each of the first $|\lsizes|$ components of a configuration $c$ is denoted by $\gamma_{cz}, z\in\lsizes$ and represents the total number of large items of size $z$. The last component, denoted by $\zeta_c$, represents the rounded cost of the bin assigned configuration $c$. Once we have the rounded cost of a bin, the number of items in a bin assigned configuration $c$ is at most ${g}^{-1}(\zeta_c)$. A configuration $c$ is feasible if the following properties hold for $c$.
\begin{enumerate}
    \item $\sum_{z\in\lsizes}z\gamma_{cz}\leq 1$,
    \item $\sum_{z\in\lsizes}\gamma_{cz}\leq g^{-1}(\zeta_c)$, and
    \item $\zeta_c \leq \costguess$.
\end{enumerate}
Let $\confs$ be the set of feasible configurations and we can compute this set in polynomial time since $|\confs|\leq (\log_{(\oeps)} f(n))\cdot(\ieps+1)^{(\ieps^3-1)}$.

\paragraph{Expensive and non-expensive configurations.}
Let $\expensivebound$ be defined as follows.
\[\expensivebound = \frac{\eps^2}{(\ieps+1)^{(\ieps^3-1)}} \hspace*{0.5em},\] and $\expensivebound$ is a constant when $\eps$ is fixed.
A configuration $c$ is \emph{expensive} if the cost of the configuration is greater than $\expensivebound\costguess$, else it is called \emph{non-expensive} (or cheap). The set of expensive configurations is denoted by $\confs_{e}$ and $\confs_{e}'=\confs\backslash\confs_{e}$ denotes the set of non-expensive configurations.

\begin{lemma}\label{integer_variables}
    The number of expensive configurations is at most a constant when $\eps$ is fixed.
\end{lemma}
\begin{proof}
    The number of possible values for the rounded cost of expensive configurations is at most $\log_{(\oeps)} 1/\expensivebound + 1$ and the number of possible values for the other components is at most $(\ieps+1)^{(\ieps^3-1)}$ since each such component can have at most $\ieps+1$ values and $|\lsizes| \leq 1/\eps^3-1$. Thus,
    \[|\confs_{e}|\leq \left(\log_{(\oeps)}1/\expensivebound + 1\right) \cdot (\ieps+1)^{(\ieps^3-1)}\]
    and it is a constant when $\eps$ is fixed.
    \qed\end{proof}

\subsubsection{Configuration MILP}
The configuration MILP we use is defined below.

\paragraph{Decision variables and constants.}
Let $n(z)$ denote the number of large items of size $z$ in the rounded instance. The decision variables are
\begin{enumerate}
    \item $v_c$ which denotes the number of bins assigned configuration $c\in\confs$, and
    \item $w_{ic}$ which denotes what fraction of small item $i$ is assigned to configuration $c\in\confs$.
\end{enumerate}
The variables in $\textbf{v}$ that correspond to expensive configurations are forced to be integer while the other variables are allowed to be fractional (in the MILP). Thus, the number of integer decision variables is at most a constant when $\eps$ is fixed from Lemma \ref{integer_variables}.

\paragraph{The MILP.}
The configuration MILP is as follows.
\begin{align}
    \min\quad
     & \sum_{c\in\confs}\zeta_cv_c                                                                                        \\
    \text{s.t.}
     & \sum_{c\in\confs}\gamma_{cz}v_c\geq n(z),\ \forall z\in\lsizes
    \label{large_cardinality_constraint}                                                                                  \\
     & \sum_{c\in\confs}w_{ic} \geq 1,\ \forall i\in\sitems
    \label{all_sitems_constraint}                                                                                         \\
     & \sum_{i\in\sitems}w_{ic} \leq \left(g^{-1}(\zeta_c) - \sum_{z\in\lsizes}\gamma_{cz}\right)v_c,\ \forall c\in\confs
    \label{confs_cardinality_constraint}                                                                                  \\
     & \sum_{i\in\sitems}s_iw_{ic} \leq \left(1 - \sum_{z\in\lsizes}z\gamma_{cz}\right)v_c,\ \forall c\in\confs
    \label{confs_size_constraint}                                                                                         \\
     & v_c\in\nre,\ \forall c\in\confs_{e}'                                                                               \\
     & v_c\in\nint,\ \forall c\in\confs_{e}                                                                               \\
     & w_{ic}\in\nre\ \forall i\in\sitems,c\in\confs
\end{align}

The motivation for this MILP is as follows (we will prove formally below that we can use the stated formulation regardless of this motivation).  The objective function is the total cost of all the configurations chosen by the MILP. The cost of each configuration is defined as part of the configuration. Constraints \eqref{large_cardinality_constraint} ensure that the chosen configurations have enough positions to pack the large items in the rounded instance. Constraints \eqref{all_sitems_constraint} ensure that all the small items are assigned to configurations. Constraints \eqref{confs_cardinality_constraint} ensure that the total number of small items assigned to a configuration, on average, is at most the cardinality left in the configuration after the large items. Constraints \eqref{confs_size_constraint} ensure that the small items assigned to a configuration, on average, is such that the total size of these small items is at most the space left in the configuration after the large items. The remaining constraints ensure the non-negativity and integrality requirements of the respective decision variables.

\begin{lemma}
    The cost of the optimal solution of the MILP is at most $(\oeps)f(\opt_\dense)$.
\end{lemma}
\begin{proof}
    We  exhibit a feasible solution to the MILP based on $\opt'_\dense$, the optimal packing of the rounded instance. Let $B$ be the set of bins in $\opt'_\dense$.

    First, we generate the configurations used by the bins in $B$. For each bin $b\in B$ do the following. Let $c(b)$ be the configuration of the bin whose components are defined as follows. Set $\zeta_{c(b)}$ equal the rounded cost of the bin. For each $z\in\lsizes$, set the $\gamma_{c(b)z}$ value to the number of large items of size $z$ in the bin. Let $\mathcal{C}$ be the multi-set of all the generated configurations. The generated configurations are feasible since $\opt'_\dense$ is a feasible packing, and we round up the cost of each bin.

    Next, we define the solution to the MILP. For each $c\in\confs$, do the following. Set $w_{ic}=1$ if small item $i$ is assigned to a bin whose generated configuration is $c$. Set $v_c$ equal to the multiplicity of $c$ in $\mathcal{C}$. Now we show the feasibility of the generated solution. Constraints \eqref{large_cardinality_constraint} and \eqref{all_sitems_constraint} are satisfied from the definition of configurations and from the feasibility of $\opt'_\dense$. Consider a bin $b$ whose generated configuration is $c(b)$. The number of small items in $b$ is equal to $|b| - \sum_{z\in\lsizes}\gamma_{c(b)z}$. Since we have a cardinality $k\in\cards'$ such that $g(k) = g(|b|) = \zeta_{c(b)}$ and $|b|\leq k = g^{-1}(\zeta_{c(b)})$, the number of small items in $b$ is at most $g^{-1}(\zeta_{c(b)})  - \sum_{z\in\lsizes}\gamma_{c(b)z}$. Considering all the copies of the configuration we see that Constraint \eqref{confs_cardinality_constraint} is satisfied. The total size of the small items in $b$ is at most $1-\sum_{z\in\lsizes}z\gamma_{c(b)}$. Considering all the copies of $c$, we see that Constraint \eqref{confs_size_constraint} is satisfied.

    From Constraint \eqref{confs_cardinality_constraint}, the number of items in a configuration chosen by the MILP will be at most the number of items in the corresponding bin. Thus, the total cost of the generated configurations is at most the total cost of the bins in $\opt'_\dense$, and the total cost of the MILP solution is, from Lemma \ref{dense_rounded_instance} and \ref{rounded_cost}, at most $(\oeps)f(\opt_\dense)$.
    \qed\end{proof}

The MILP has a polynomial number of constraints and variables. The number of variables forced to be integers is at most a constant once $\eps$ is fixed. Thus, the MILP can be solved in polynomial time using the approach of \cite{kannan1983improved,lenstra1983integer}.

Consider the solution to the MILP denoted as  $(\textbf{v}, \textbf{w})$ where $v_c, \forall c\in\confs_e$ is an integer. Round up the $\textbf{v}$ to $\textbf{v}'$ i.e., $v'_c = \lceil v_c\rceil$ for all configurations $c$. We add at most one extra bin for each non-expensive configuration $c$, and such added bins are called \emph{supplementary bins}. Next, we bound the total cost of these supplementary bins.

\begin{lemma}\label{cost_supplementary}
    The total cost of the supplementary bins is at most $4\eps f(\opt_\dense)$.
\end{lemma}
\begin{proof}
    The number of non-expensive configurations for a fixed rounded cost $k$ is at most \\ $(\ieps+1)^{(\ieps^3-1)}$. Thus, the total cost for all supplementary bins of rounded cost $k$ is at most $(\ieps+1)^{(\ieps^3-1)}\cdot k$. Since the largest cost of a supplementary bin is $\expensivebound\costguess$ and since the cost function was geometrically rounded, the total cost of all supplementary bins is at most
    \[\sum_{l=0}^{\infty}(\ieps+1)^{(\ieps^3-1)} \frac{\expensivebound\costguess}{(\oeps)^l}\hspace{0.5em}.\]
    Using the definition of $\expensivebound$ and Lemma \ref{rounded_cost},
    \[\sum_{l=0}^{\infty}(\ieps+1)^{(\ieps^3-1)} \frac{\expensivebound\costguess}{(\oeps)^l} \leq (\ieps+1)^{(\ieps^3-1)} \frac{(\oeps)\expensivebound\costguess}{\eps}
        \leq 2\eps\costguess\leq 2\eps g(\opt_\dense) \leq 4\eps f(\opt_\dense).\]
    \qed\end{proof}

\subsubsection{Packing the rounded instance based on MILP solution}
The packing of the large items is standard based on the configuration counters while for packing the small items we are using a method that we previously used in \cite{jaykrishnan2024scheduling}.

For each $c$ in the support of $\textbf{v}$, assign configuration $c$ to $v'_c$ bins. Let $\bins$ be the set of bins we have after rounding up the MILP solution including the supplementary bins (that is, $\sum_{c} v'_c$ bins) and let $\bins_c$ be the set of bins whose assigned configuration is $c$.

Pack the large items to all the bins in $\bins$ according to the configuration corresponding to the bins. We have more positions for large items than necessary and so some positions for large items may be empty.  So if a configuration $c$ is assigned to a bin $b$, we allow up to $\gamma_{cz}$ items of size $z$ for every size $z$ of large items.

The packing of the small items is done as follows. For each bin $b$ in $\bins$ with configuration $c(b)$, calculate the space available for packing small items in $b$ as $\rho_b = 1 - \sum_{z\in\lsizes}z\gamma_{c(b)z}$, and the maximum number of small items that can be packed in $b$ as $\omega_b = g^{-1}(\zeta_{c(b)}) - \sum_{z\in\lsizes}\gamma_{c(b)z}$.  We use the following feasibility linear program LP to get a feasible fractional assignment of small items to the bins such that the assignment satisfies the space and cardinality constraint for each bin. Here $\mu_{ib}$ is a decision variable for each small item $i$ and bin $b\in\bins$.
\begin{align}
    \sum_{i\in\sitems} \mu_{ib}    & \leq \omega_{b}, \forall b\in\bins      \label{LP_bin_cardinality_constraint}           \\
    \sum_{i\in\sitems} s_i\mu_{ib} & \leq \rho_b, \forall b\in\bins       \label{LP_bin_space_constarint}                    \\
    \sum_{b\in\bins} \mu_{ib}      & = 1, \forall i\in\sitems                                     \label{LP_all_small_items} \\
    \mu_{ib}                       & \geq 0, \forall i\in\sitems, \forall b\in\bins
\end{align}

\begin{lemma}
    LP has a feasible solution.
\end{lemma}
\begin{proof}
    The MILP solution, after rounding, is $(\textbf{v}', \textbf{w})$. From Constraints \eqref{confs_cardinality_constraint} and \eqref{confs_size_constraint} of the MILP, and the rounding of the MILP solution, we have the following.
    \begin{align}
         & \sum_{i\in\sitems}\frac{w_{ic}}{v'_c}
        \leq \left(g^{-1}(\zeta_c) - \sum_{z\in\lsizes}\gamma_{cz}\right) \frac{v_{c}}{v'_c}
        \leq \left(g^{-1}(\zeta_c) - \sum_{z\in\lsizes}\gamma_{cz}\right). \label{inequality_1} \\
         & \sum_{i\in\sitems}s_i\frac{w_{ic}}{v'_c}
        \leq \left(1 - \sum_{z\in\lsizes}z\gamma_{cz}\right)\frac{v_{c}}{v'_c}
        \leq \left(1 - \sum_{z\in\lsizes}z\gamma_{cz}\right) \label{inequality_2}.
    \end{align}
    The above inequalities are true for a configuration $c$, consequently the inequalities also apply for a bin assigned configuration $c$. Set $\mu_{ib}$ equal to ${w_{ic(b)}}/{v'_{c(b)}}$ where $c(b)$ is the configuration assigned to bin $b$ for all $i\in\sitems$. From the inequalities \eqref{inequality_1} and \eqref{inequality_2}, we see that $\mu_{ib}={w_{ic(b)}}/{v'_c(b)}$ satisfies constraints \eqref{LP_bin_cardinality_constraint} and \eqref{LP_bin_space_constarint} of the LP.
    Let $\bins_c$ be the set of bins whose assigned configuration is $c$ and $|\bins_c|=v'_c$. Notice that, $\sum_{b\in\bins}\mu_{ib} = \sum_{c\in\confs}\sum_{b\in\bins_c}\mu_{ib} = \sum_{c\in\confs}w_{ic}$. From Constraint \eqref{all_sitems_constraint} of the MILP, $\sum_{b\in\bins}\mu_{ib} \geq 1$ and $\mu_{ib}={w_{ic(b)}}/{v'_c(b)}$ satisfies Constraint \eqref{LP_all_small_items} of the LP as inequality. If it is a strict inequality, we can safely decrease the $\mu$ vector so that it will be satisfied as equality without hurting the other inequalities.  Thus, there is a feasible solution to the LP.
    \qed\end{proof}

Let $\mu^*$ be a basic feasible solution. The number of strictly positive components in $\mu^*$ is at most $2|\bins|+|\sitems|$. If there is only one positive component in the $\mu^*$ values for a small item, then that component must be equal to $1$. Thus, the number of items $i$ such that the sub-vector of $\mu^*$ corresponding to item $i$ is not integral is at most $2|\bins|$. Remove at most $2|\bins|$ small items, whose $\mu^*$ sub-vector is fractional, and pack the remaining small items to bins according to the $\mu^*$ values. Pack the removed small items into at most $2\eps|\bins|+1$ bins so that every such bin has at most $1/\eps$ small items.

\begin{lemma}\label{dense_small_items_cost}
    The increase in the rounded cost due to these additional bins is at most $2\eps$ times the cost of the rounded MILP solution $(\textbf{v}', \textbf{w})$ plus $g(\ieps)$.
\end{lemma}
\begin{proof}
    The number of new additional bins required is at most $2\eps|\bins|+1$ and the rounded cost of each such bin is at most $g(\ieps)$. Since we are dealing with the dense bins, the cost of any configuration of a dense bin is at least $g(\ieps)$. Thus, the cost of the additional bins is at most $2\eps$ times the cost of the rounded  solution  $(\textbf{v}', \textbf{w})$ plus an additive term of $g(\ieps)$ as claimed.
    \qed\end{proof}

Thus, we conclude the following theorem.

\begin{theorem}
    $\gcbp$ admits an APTAS.
\end{theorem}

\bibliographystyle{abbrv}

\begin{thebibliography}{10}

\bibitem{anily1994worst}
S.~Anily, J.~Bramel, and D.~Simchi-Levi.
\newblock Worst-case analysis of heuristics for the bin packing problem with
  general cost structures.
\newblock {\em Operations research}, 42(2):287--298, 1994.

\bibitem{bramel1997average}
J.~Bramel, W.~T. Rhee, and D.~Simchi-Levi.
\newblock Average-case analysis of the bin-packing problem with general cost
  structures.
\newblock {\em Naval Research Logistics (NRL)}, 44(7):673--686, 1997.

\bibitem{caprara2003approximation}
A.~Caprara, H.~Kellerer, and U.~Pferschy.
\newblock Approximation schemes for ordered vector packing problems.
\newblock {\em Naval Research Logistics (NRL)}, 50(1):58--69, 2003.

\bibitem{epstein2010afptas}
L.~Epstein and A.~Levin.
\newblock {AFPTAS} results for common variants of bin packing: A new method for
  handling the small items.
\newblock {\em SIAM Journal on Optimization}, 20(6):3121--3145, 2010.

\bibitem{epstein2012bin}
L.~Epstein and A.~Levin.
\newblock Bin packing with general cost structures.
\newblock {\em Mathematical programming}, 132:355--391, 2012.

\bibitem{epstein2017afptas}
L.~Epstein and A.~Levin.
\newblock An {AFPTAS} for variable sized bin packing with general activation
  costs.
\newblock {\em Journal of Computer and System Sciences}, 84:79--96, 2017.

\bibitem{fernandez1981bin}
W.~Fernandez~de La~Vega and G.~S. Lueker.
\newblock Bin packing can be solved within 1+ $\varepsilon$ in linear time.
\newblock {\em Combinatorica}, 1(4):349--355, 1981.

\bibitem{haouari2024lower}
M.~Haouari and M.~Mhiri.
\newblock Lower and upper bounding procedures for the bin packing problem with
  concave loading cost.
\newblock {\em European Journal of Operational Research}, 312(1):56--69, 2024.

\bibitem{jansen2019approximation}
K.~Jansen, M.~Maack, and M.~Rau.
\newblock Approximation schemes for machine scheduling with resource (in-)
  dependent processing times.
\newblock {\em ACM Transactions on Algorithms (TALG)}, 15(3):1--28, 2019.

\bibitem{jaykrishnan2024scheduling}
G.~Jaykrishnan and A.~Levin.
\newblock Scheduling with cardinality dependent unavailability periods.
\newblock {\em European Journal of Operational Research}, 2024.

\bibitem{kannan1983improved}
R.~Kannan.
\newblock Improved algorithms for integer programming and related lattice
  problems.
\newblock In {\em Proceedings of the fifteenth annual ACM symposium on Theory
  of computing}, pages 193--206, 1983.

\bibitem{karmarkar1982efficient}
N.~Karmarkar and R.~M. Karp.
\newblock An efficient approximation scheme for the one-dimensional bin-packing
  problem.
\newblock In {\em 23rd Annual Symposium on Foundations of Computer Science
  ({FOCS} 1982)}, pages 312--320. IEEE, 1982.

\bibitem{kellerer1999cardinality}
H.~Kellerer and U.~Pferschy.
\newblock Cardinality constrained bin-packing problems.
\newblock {\em Annals of Operations Research}, 92(0):335--348, 1999.

\bibitem{lenstra1983integer}
H.~W. Lenstra~Jr.
\newblock Integer programming with a fixed number of variables.
\newblock {\em Mathematics of operations research}, 8(4):538--548, 1983.

\bibitem{leung2008asymptotic}
J.~Y.-T. Leung and C.-L. Li.
\newblock An asymptotic approximation scheme for the concave cost bin packing
  problem.
\newblock {\em European journal of operational research}, 191(2):582--586,
  2008.

\bibitem{li2006bin}
C.-L. Li and Z.-L. Chen.
\newblock Bin-packing problem with concave costs of bin utilization.
\newblock {\em Naval Research Logistics (NRL)}, 53(4):298--308, 2006.

\bibitem{vazirani2001approximation}
V.~V. Vazirani.
\newblock {\em Approximation algorithms}, volume~1.
\newblock Springer, 2001.

\end{thebibliography}

\end{document}